\documentclass[12pt,preprint]{aastex}

\shorttitle{IGR J17464-3213/H1743-322 state transition and flaring activity}
\shortauthors{Joinet et al.}

\begin{document}
\title{State transition and flaring activity of
 IGR J17464-3213/H1743-322 with SPI/INTEGRAL telescope}
\author{A. Joinet\altaffilmark{1}, E. Jourdain\altaffilmark{1}, J. Malzac\altaffilmark{1}, J.P. Roques\altaffilmark{1}, V.Sch\"{o}nfelder\altaffilmark{2}, P. Ubertini\altaffilmark{3}, F. Capitanio\altaffilmark{3}}
%\affil{CESR, 9 Avenue du Colonel Roche, BP4346, 31028 Toulouse cedex , France}
%\affil{National Optical Astronomy Observatories, Tucson, AZ 85719}
%\email{ubertini@rm.iasf.cnr.it}
%
%\and
%
%\author{P. Ubertini\altaffilmark{2}, F. Capitanio\altaffilmark{2}}
%\affil{Space Telescope Science Institute, Baltimore, MD 21218}
\altaffiltext{1}{CESR, 9 Avenue du Colonel Roche, BP4346, 31028 Toulouse, France}
\altaffiltext{2}{Max-Planck-Institut f\"{u}r Extraterrestrische Physik, Postfach 1603, 85740 Garching, Germany} 
\altaffiltext{3}{IASF-CNR, via del Fosso del Cavaliere 100, 00133 Roma,Italy} 
%\altaffiltext{3}{}
%\altaffiltext{4}{Visiting Programmer, Space Telescope Science Institute}
%\altaffiltext{5}{Patron, Alonso's Bar and Grill}
%
%

\begin{abstract}
IGR J17464-3213, already known as the HEAO-1 transient source  H1743-322,
 has been detected during a state transition by the SPI/INTEGRAL telescope. 
We describe the spectral evolution and flaring activity of IGR J17464-3213/H1743-322 
 from 2003 March 21 to 2003 April 22. During the first part, the source followed a continuous
spectral softening, with the peak of  the spectral energy distribution shifting
  from 100 keV down to $\sim$ a few keV.
%The lack of an X-ray blackbody 
% attention lack: of X-ray blackbody component= hard state !!!
 However the thermal disk and the hard X-ray components had a similar intensity,
 indicating that the source was in an 
  intermediate state throughout our observations and evolving toward the soft state.
  In the second part of our observations, the ASM/RXTE and SPI/INTEGRAL light curve showed
   a strong flaring activity. 
  %while the
  %broad band spectral shape shows only small changes.
  Two flare events lasting about 1 day each have been detected with SPI and are probably due to 
  instabilities in the accretion disk associated with the state transition.
  During these flares, the low  (1.5-12 keV) and high (20-200 keV) energy fluxes monitored with 
  ASM/RXTE  and SPI/INTEGRAL, 
 are correlated and the spectral shape (above 20 keV) remains unchanged while the luminosity increases by
  a factor greater than 2.

\end{abstract}
\section{Introduction}
On 2003 March 21, during a scan of the Galactic center 
region, the new INTEGRAL source  IGR J17464-3213  was detected with 
the IBIS imager \citep{REV}.
Following this INTEGRAL observation,  RXTE observed 
the same field and found  XTE J1746-322 at a 
 position consistent with that of IGR J17464-3213  \citep{Mark}
 with a hard spectrum (with a powerlaw photon index $\Gamma$ of $\simeq 1.49 \pm 0.01$), while a radio 
 counterpart  has been identified with the VLA \citep{Rupa}. 
All the positions  were compatible \citep{Rem}
with the  H1743-322  transient source discovered  during its outburst in 1977
by Ariel V and HEAO-1 \citep{Kaluza} which had a maximum observed flux 
 of 730 mCrab \citep{Kaluzb}.\\

\begin{figure*}%[!t]
\begin{center}
\includegraphics[scale=0.8]{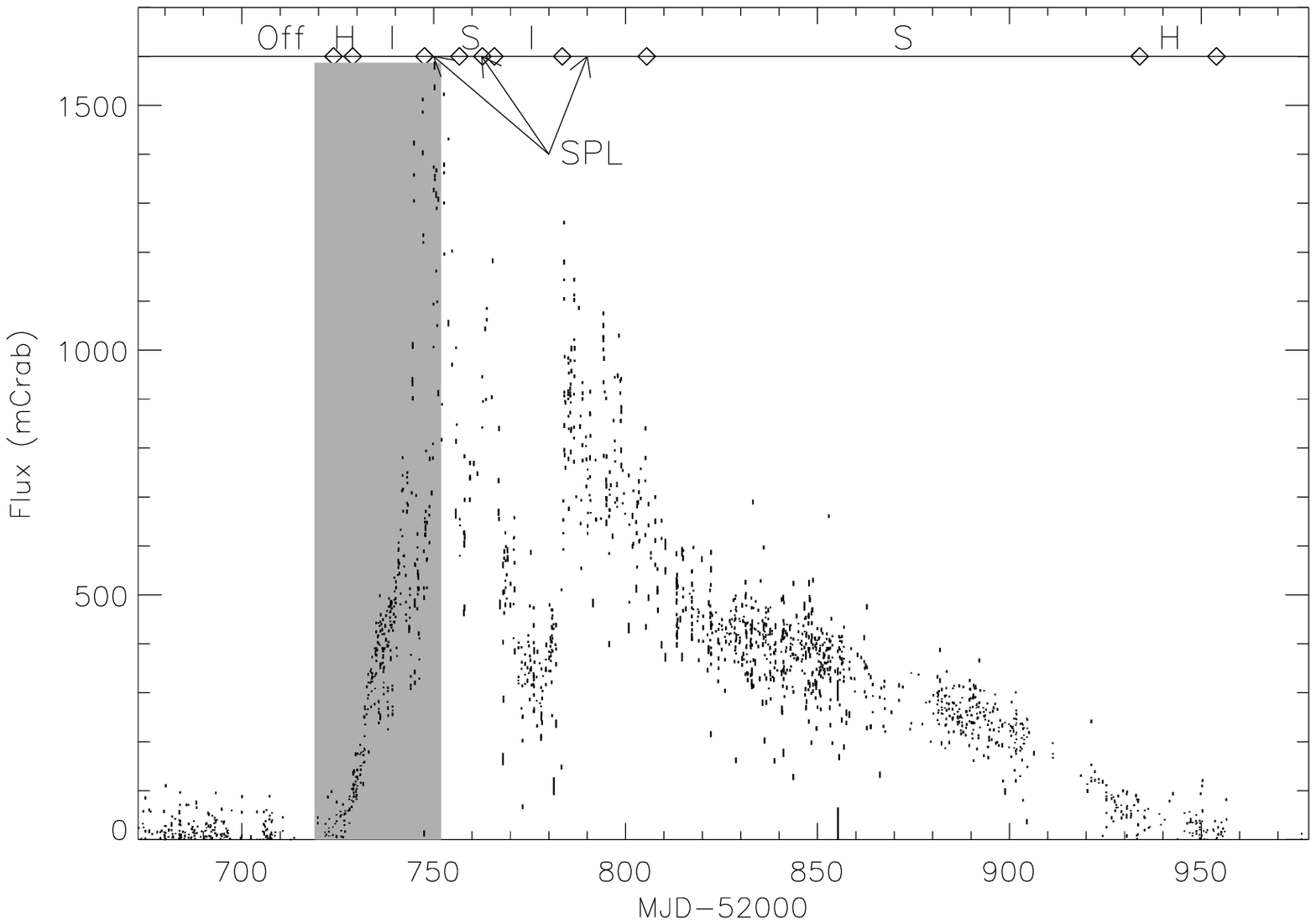}
\end{center}
%\hspace{2cm} 
%\vspace{-0.5cm}
%\vspace{-0.4cm}
\caption{ASM/RXTE light curve (1.5-12 keV) covering the 2003 outburst of H1743-322.  The different states summarized on the graph fulfill the
 criteria (see the text) of the spectral classification of McClintock \& Remillard (2003). The corresponding legend is:
  Off : off, H: hard, I: intermediate, S: thermal dominant and SPL. 
  The grey color bar shows the ASM/RXTE observations
 for which the SPI/INTEGRAL data have been analyzed. Figure 3 (see below) focuses on these observations.} 
\label{fig:figA}
\end{figure*}

Early 2003,  INTEGRAL observations showed a slowly rising hard X-ray flux.
 That was the beginning of a long outburst which lasted until 
 the end of November 2003. 
 During this period, the source was  monitored by  
 RXTE (Homan et al. 2004; Remillard et al. 2004) and  Chandra (Miller et al. 2004)
 showing the occurrence of various spectral/timing states. In  some of them it showed
  high frequency quasi-periodic oscillations 
  at 240 Hz and 160 Hz (therefore in the 3:2 ratio which has 
 recently emerged in a few Galactic black hole systems) compatible with the dynamical time-scale
  at the innermost stable orbit around a 10 solar mass black hole. 
  The presence of dips in the X-ray light curves suggests a large inclination 
  ($\sim$ 60-70$\degr$) (Homan et al. 2004).  
 Observations with the VLA showed intense radio activity,
  in particular a strong radio flare was reported on April 8
   associated with a jet \citep{Rupb}.
  Infrared  \citep{baba}
 and optical \citep{steeghs} counterparts were reported  and found to be consistent 
 with emission from a jet. 
 The ASM/RXTE light curve of the 2003 outburst is presented in Fig.~\ref{fig:figA}. 
 Its profile is more complex than the fast rise slow decay of classical X-ray novae, 
 but shows some similarities with other transient black hole X-ray binaries in outburst
  such as XTE J1859+226 \citep{Broc}, GRO J1655-40 \citep{Rema} 
  or XTE J1550-564 \citep{Sob}.

 Due to the source location near the Galactic Center, further INTEGRAL observations of 
 IGR J17464-3213 
  were carried out in the framework of the Galactic Centre Deep Exposure (GCDE) 
  of the  core program \citep{Winkl} in March, April, August and September 2003
    (Parmar et al. 2003; Lutovinov et al. 2004; Capitanio et al. 2005).
    While these monitoring campaigns may suggest a 
    smooth evolution from the hard to the soft state, RXTE
    observations, together with a bigger set of data, allow us to study
    more in detail the single phases of the source
    evolution during the rising part of the outburst.The
    timing results, which can be seen from Table 1 in the H17433-322 paper by Remillard et al. (2004) together
    with the overall X-ray colour evolution,
     suggest, refering to the McClintock \& Remillard (2003) classification,
         the following spectral state evolution :  "off" $\rightarrow$ hard $\rightarrow$ intermediate $\rightarrow$ SPL $\rightarrow$ thermal dominant 
 $\rightarrow$ SPL $\rightarrow$ intermediate $\rightarrow$ SPL 
 $\rightarrow$ thermal dominant $\rightarrow$ hard  
 $\rightarrow$ "off" (shown on Fig.~\ref{fig:figA}).

 To summarize, the timing and spectral properties of the source
 indicate a low mass X-ray binary harboring a black hole 
  observed at large inclination (Homan et al. 2004; Lutovinov et al. 2004).

 We  present here the results of the SPI instrument observations on board 
INTEGRAL together with PCA/RXTE data for the X-ray emissions, during Spring 2003,  at the beginning of the outburst when the source was 
evolving through an intermediate state. 
%in X-rays (1.5-12 keV, ASM data)
%, probably representing  the transition
%from a low/hard to a high/soft state similar to that seen in  
%Black Hole candidates.

Some IBIS and JEM-X  data obtained during the same observations 
have already been published in Parmar et al. (2003) and Capitanio et al. (2005).
 %The source  shows, above 15 keV 
%(ISGRI/INTEGRAL data), 
%a slight hardening of the spectrum as the
%intensity increases, which is opposite to the softening presented in the X-ray  band 
%(ASM /RXTE data). The INTEGRAL data showed
%the evolution of the source  geometry from the hard 
%comptonised emission with  the appearance of a soft disk component. 
%{\bf en fait la soft component apparait pendant l'ete et pas pendant les obs de 
%ce  papier, non ?}
%Radio observations showed a strong radio flare during 
%on April 8 (Rupen et al. 2003b).\\ 
Here we use a much larger set of SPI data (all the publicly available data)
to study in more detail the spectral evolution 
 of  H1743-322.

\section{ SPI/INTEGRAL instrument}
SPI (Spectrometer for Integral, Vedrenne et al. 2003) is one of INTEGRAL's two main 
instruments. Working in the 20 keV - 8 MeV energy 
domain with 19 hexagonal germanium detectors, it possesses an excellent energy 
resolution on a 16 degree (corner to corner) field of view. \\
Careful calibration performed before launch using radioactive
 sources \citep{Att} and extensive modelling of the instrument led to a
precise determination of its response. The imaging and spectral performance 
has since been verified in flight \citep{Roq}. The knowledge of the SPI
instrument along with its spectroscopic capabilities allows us to obtain precise results
for the spectral behaviour of X/$\gamma$- ray sources. 

SPI's imaging
capability is limited but it reaches a $2.5^{\circ}$ angular resolution thanks
to a HURA (Hexagonal Uniform Redundant Array) coded mask, whose cells have 
the same size as the individual detectors. Due to the small number of 
detectors (pixels), SPI image reconstruction methods are based on a combination
of data from several pointings separated by 2 degrees and covering
the same sky region ("dithering strategy", see Jensen et al. 2003).

During the  $\simeq$3 day INTEGRAL revolution, the observing schedule consists 
of fixed pointings lasting approximatively 30-40 minutes, with a complete 
dithering pattern made up of 7 (in hexagonal mode) or 25 
pointings (in rectangular mode) separated by a 2 $\degr$ angular
distance. The latter one is the most commonly used mode.
This method increases the amount of data  in excess of the  number
of unknowns and allow to better determine the background and 
the position and the flux of sources on the detector plane. 

\section{Data}
The analysis presented here is based on the data recorded from revolution 53 
starting 2003 March 21 to revolution 63 finishing 2003 April 22 
(it corresponds to MJD 52719-52751).\\
Some pointings had to be removed due to several reasons. For strong sources, 
if the projected area onto the detection plane and/or the number of pointings 
are insufficient, it leads to a lack of sensitivity and/or 
number of equations to solve (1). Consequently the source flux is determined 
with huge error bars. Sco~X-1 and in certain cases 4U~1700-377 
appear in a too small part of a dithering pattern, leading to artefacts 
during the image  reconstruction process. We excluded such pointings and also those 
affected by a solar flare or by 
exit/entry into the radiation belts. Otherwise, the background is generally stable. 
Table~\ref{tab:table1} gives the details of each revolution. 
The 230 pointings used for the SPI data analysis give a useful 
exposure time of $\sim$333 ks. 

Public PCA/RXTE data, covering some part of the SPI/INTEGRAL data observation periods,
were available. Table~\ref{tab:table1} summarizes the set of PCA/RXTE observations
performed under the program ID 80138 and 80146.

\section{Analysis method}
\subsection{SPI data analysis}
The signal recorded by the SPI camera on the 19 Ge detectors is composed of 
the contribution from each source in the
 field of view through the instrument aperture plus the background, which 
 comes mainly from the interaction of high energy particle (
 coming from cosmic ray or due to solar activity) with the instrument.\\
For $N_s$ sources present in the field of view,  
the data $D_{p}$ obtained during a pointing $p$
for a given energy band, can be expressed by the relation:

\begin{equation}
  D_{p} = \sum_{i=1}^{N_s} {R_{p,i} \times S_{p,i} } + B_{p} 
\end{equation}

where $R_{p,i}$ is the response of the instrument for the source $i$,
$S_{p,i}$ the flux of the source  $i$  and $B_{p}$ the background
recorded during the pointing $p$. $D_{p}, R_{p,i}$ and $B_{p}$ are vectors of 19 elements.\\
 It is mandatory  to reduce the number of unknowns related to the background.
 In the present work, we describe the background as  $ B_{p} = A_{p} U $ where  $A_{p}$ is 
  the normalization coefficient per pointing and U the "uniformity map" of the SPI camera. 
 This uniformity map is derived from an empty field observation.
$N_{d} \times N_{p}$
The system consists of $N_{d}$ (number of detectors) $\times$ $N_{p}$ (number of pointings) equations
 solved simultaneously by a chi-squared minimisation method.
 The number of unknowns (free parameters) is $N_{p}$ $\times$ ($N_s$ + 1) (for the $N_s$ sources and 
 the background fluxes)
  but we can limit
 them by assuming that the time variability for the sources and the background is longer than a 
 single pointing.\\ 
The  timescales are chosen depending on the goal of the analysis. For the image   reconstruction, performed by the SPIROS 
detection algorithm \citep{Ski}, source
 and background  fluxes are assumed to be constant during the whole set of observations.
 Using an iterative source research technique, implemented 
for the coded mask telescope SPI/INTEGRAL, the
source positions are extracted. The fluxes determined in this way are thus rough mean fluxes,
the main goal being to extract source positions.   

When wishing to build the light curves of the sources detected in the field of view, we 
must choose the appropriate time scales for each component (source and background).\\
Concerning the background flux, the global count-rates registered with the Anti-Coincidence System (ACS) 
indicate  that the background flux evolves within a single revolution. 
A timescale of 6 hours seemed to be the best value to describe the background variability 
during our observations.
For the point-like sources, it is important not to oversample the temporal variability as 
this increases the error bars and gives no further scientific information.
 We have chosen a time scale for each source
mainly as a function  of its intensity and of its temporal behavior.   The fainter sources 
 have been considered to have a constant flux within a revolution. 
   For the brightest sources, we test several values and choose the longest timescale
   over which the source is found not to vary (see below). 

 Spline functions are used in order to divide one interval of observations 
(one revolution or part of it) into several subintervals ($\Delta t_i \geq$ pointing duration) 
during which 
the flux of the i-th component (source or background) is constant.

The  system of Np*Nd equations (1) is thus completed by a number of additional constraints reflecting
the non-variability of a given parameter ($S_{i,p}$ or $A_{p}$) over its timescale ($\Delta t_i$).
Its resolution by the chi-squared minimisation method gives the light curves of all components simultaneously.

The count spectra are constructed by solving a similar system in a number of energy bands and then
deconvolved using  the energy response matrix corresponding to each pointing \citep{Stu}  
to get the photon spectra. 

\subsection{RXTE Spectral reduction}
We analysed the PCA/RXTE (Proportional Counter Array) (Bradt et al. 1993) data.
The data reduction in the 3-20 keV energy range were performed using the "ftools" routines in
the HEAsoft software package distributed by NASA's HEASARC.  
The spectrum extraction was performed from data taken in "Standard 2" mode. The response matrix and 
the background model were created using ftools 
programs. Background spectra were made using the latest "bright source" background model.
We used all available PCUs for the first part of observations.
Since  PCU$\#$2 was the  detector always in use during the observations from rev. 60 up to 63,
it has been used for the data extraction in order to sum several spectra. 
We added 0.8\% up to 7 keV and 0.4\% above 7 keV as 
systematic error.

\section{Results}
\subsection{Images}
\begin{figure*}%[!t]
\begin{center}
\includegraphics[scale=0.8]{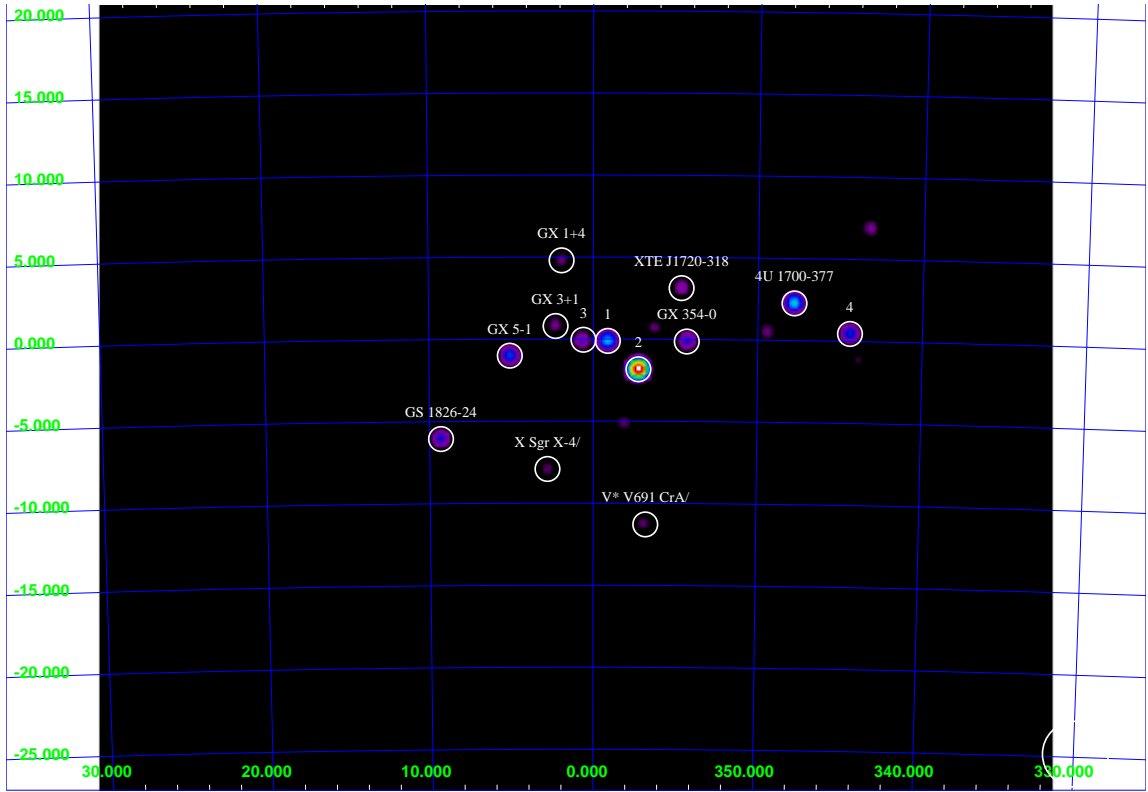}
\end{center}
%\hspace{2cm} 
%\vspace{-0.5cm}
%\vspace{-0.4cm}
\caption{SPIROS image obtained using  revolutions 56, 58, 59, 60, 61, 62 and 63 in the 20-36 keV  
energy range with a significance higher than 5 $\sigma$. The horizontal axis corresponds to the Galactic longitude
and the vertical one to the Galactic latittude. The number corresponding to 
soures are 1 : 1E 1740.7-2942, 2 : H1743-322, 3 : IGR J17475-2822, 4 : OAO 1657-415. }  
\label{fig:fig1}
\end{figure*}

The SPI images were built with the SPIROS detection algorithm \citep{Ski}. 
 We used a catalogue of known sources with their theoretical positions and searched 
 for 15 sources in this catalogue plus 5 possible new sources. 
The image of the sky observed by SPI in the 20-36 keV energy range in the 
crowed region of the Galactic
 Centre can be seen in Figure~\ref{fig:fig1}, for where 
 H1743-322 is in the field of view.
 H1743-322 is the strongest source (247 mCrab) with a significance
  of  127 $\sigma$ during the total exposure time. H1743-322 is relatively isolated since 
there are no detected sources within a distance
  of 2.5 degrees. Table~\ref{tab:table2} summarizes the 13 sources 
  surrounding H1743-322  in the 20-36 keV energy range.
  Beyond 90 keV, there are only 3 sources having a significance greater than 5 $\sigma$ and H1743-322
  is the most significant one.   

\subsection{Light curves}
\begin{figure}%[!t]
\begin{center}
\includegraphics[scale=0.9]{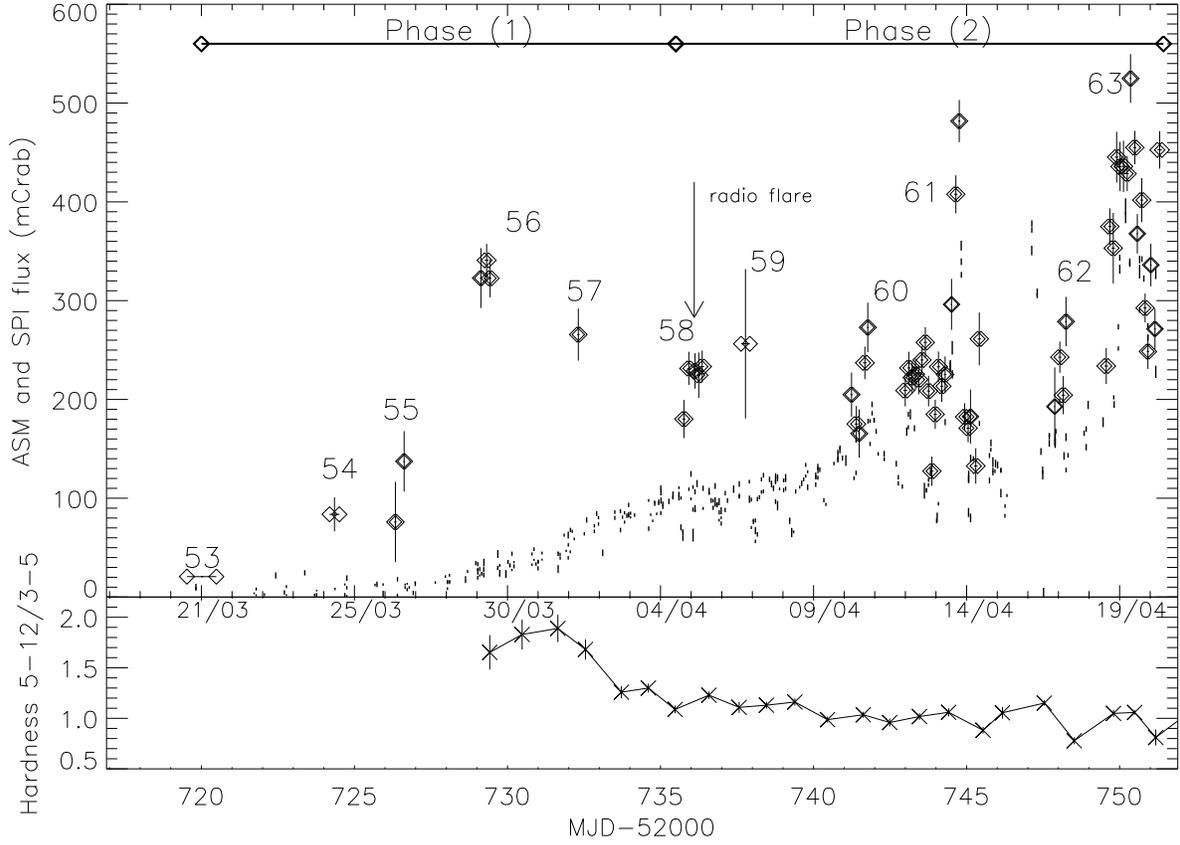}
\end{center}
%\hspace{2cm} 
%\vspace{-0.5cm}
\caption{SPI light curve  (diamonds) in the 20-36 keV energy range and  ASM (thick bar) in the 1.5-12 keV energy range.
The ASM flux has been divided by a factor of 4. We also
give the revolution number and the period covered by each state of the source.
An upper limit has been determined for revolution 53. The  panel 
below the light curve shows the evolution of the hardness from the ASM/RXTE in the 3-12 keV energy range.
A vertical arrow indicates the radio flare event (see section 1.).}
\label{fig:fig2}
\end{figure}

The light curve of H1743-322 (Figure~\ref{fig:fig2}) in the 20-36 keV energy range
 was constructed using the method described in Section 3, taking into account the 
 14  sources identified in the field of view  (Table~\ref{tab:table2}). We have kept
 the flux of fainter sources constant and fixed timescales for the 3 more intense  
 on the basis of their variability and of their flux intensity: of 7200 seconds for H1743-322,
  86400 seconds for 1E 1740.7-2942 and  10800 seconds for 4U 1700-377 which is known 
  to be variable.
  The SPI light curve is similar to that published in Capitanio et al. (2005) from IBIS data, 
  except
  for the 
 bursts around MJD 52743 and 52749 for which we added new released data to complete our analysis.
 In order to extend the spectral coverage towards lower energy, we 
extracted the light curve obtained by ASM/RXTE in 
the 2-12 keV domain in the same period of observation
(data taken from the public XTE database:  http://xte.mit.edu/lcextrct/asmsel.html). \\
Since the beginning of the RXTE measurements in 1996, until the beginning of our observations 
(i.e. 7 years) the source was remaining in
an "off" state. Then its intensity increases
rapidly over a month culminating in a huge outburst which spans several months with a
maximum at MJD 52753.

  We see a clear variability of the source on various timescales with  initially
 an enormous flux increase over
 3 days. From the "off" state (with an upper limit at 2 $\sigma$ of 20 mCrab at MJD 52719), the 20-36 keV flux  increases by a factor of 3.7 
 between MJD 52724-52727 (89  $\pm$ 13 mCrab) and MJD 52729 ( 328 $\pm$11 mCrab). 
  Meanwhile, the ASM flux increases up to $\sim$ 200 mCrab by the same factor.

  Then, while the SPI fluxes 
decrease from MJD 52729 to MJD 52742, the ASM flux follows an opposite behavior. 
Note that a radio flare occurs around 
  revolution 58 (MJD 52736) with a radio flux increase of a factor 5 between
   April 6th and 8th \citep{Rupb} (see Figure 3). 

 In a second phase, the both emissions 
   are dominated by flaring activity with $\sim$1 day 
   bursts (MJD 52744.6 in revolution 61 and MJD 52749.7 in revolution 63) during which the  flux is  typically 
   multiplied by 2 or 3. 
   Although the long timescale trends are opposite, 
   the soft X-ray and hard X-ray emission are perfectly 
   correlated inside these flares.

% the source intensity was below the detection threshold and then increases 
% rapidly over a month culminating in a huge outburst which spans several
% months  with a maximum at MJD 52753.9. 
%The SPI and ASM light curve shows an important spectral variability. 
%By analogy with other hard X-ray sources, we recognize a transition from a hard to a soft
%state, beginning around IJD 1185.
% Note that the radio flare occurs around revolution 58
%with the flux multiplied by 5 between April 6th and 8th \citep{Rupb}.\\ 
%Comparing more precisely the temporal evolution in SPI  and ASM bands, we see that, after an "off" state ...
%which is the same factor increase seen in 
 %the ASM flux. 
 %Then, during the rising part of the X-ray outburst, the mean hard X-ray  flux  slowly decreases

We can thus conclude that two distinct time scale modes contribute to the observed light curves:
one 
on a weeks timescale, which we will relate to a state transition (see below), the other mode is on a day timescale and produces 
simultaneously soft and hard X-rays. This behavior is reminiscent of 
Cygnus X-1 one, recent
observation showing a basically constant spectral shape on short time scale ($\simeq$ hours) flux
 variability \citep{Baz}.
 
\subsection{Hardness evolution}
In order to study the spectral evolution of the source, we have examined the hardness ratios 
in the 20-90 keV energy range, defined as $H=\frac{Counts{(36-90 keV)}}{Counts{(20-36 keV)}}$.
 All the fluxes  are expressed in mCrab to allow comparison with other instruments.

\begin{figure*}%[!t]
%\begin{center}
%\centering
\includegraphics[scale=0.7]{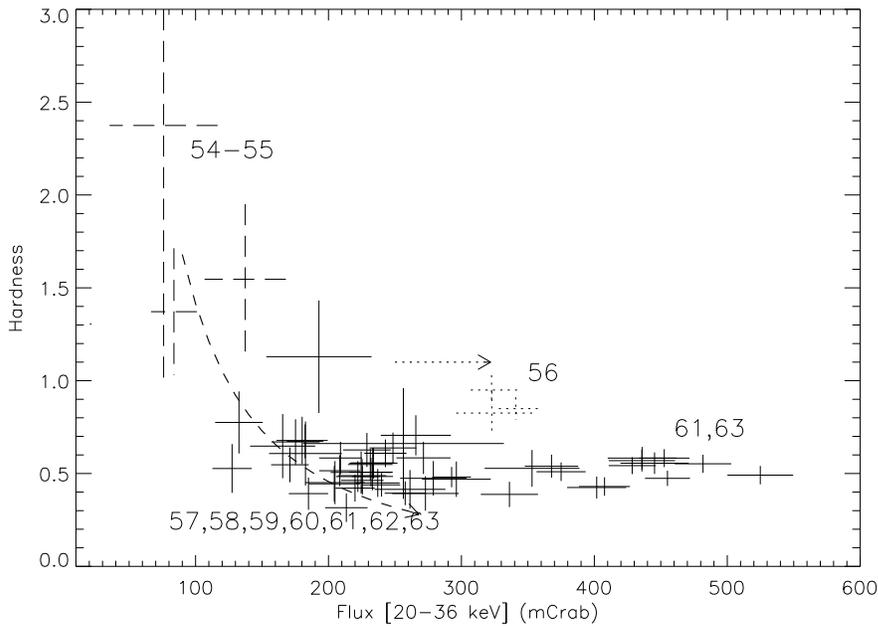}
%\end{center}
%\hspace{2cm} 
%\vspace{-0.5cm}
\caption{Overview of the hardness behaviour of H1743-322: in the hard state (revolutions 54-55 and 56, which
are represented by dashed and dotted line respectively) 
and the intermediate/SPL state ($\geq$ revolution 57). The horizontal and dotted arrow illustrates the increase of flux with a constant hardness 
from revolution 54-55 to revolution 56 (see section 5.3). A curved dashed line has been drawn to illustrate the negative
 hardness-flux correlation in the 20-36 keV energy range.}  
\label{fig:hardtot}
\end{figure*}

Figure~\ref{fig:hardtot} shows the evolution of the hardness for revolutions 54 to 63.
We clearly see 2 groups of points, with mean fluxes of $\sim$220 and $\sim$460 mCrab 
respectively. During the low level flux periods,
there is a clear  hardness-flux anti-correlation 
with a linear correlation factor of $\simeq$ - 0.7.
It shows clearly the transition from 
 the hard state to a softer  state on a timescale of $\sim$2 weeks.
Conversely, the flux increases at constant hardness during the flare events of revolutions 61 and 63.
We also note that  data from revolution 56 are not in the continuity of the revolutions 54-55 and 57.
But they could correspond to a flux increase with a constant hardness 
(see the horizontal arrow in Fig. 4) relatively to 
the anticorrelation trend.

%Figure~\ref{fig:hard54-58} shows the evolution of the hardness over a few hours
%for revolutions 54  to 58. 
%  There is a clear  hardness-flux anti-correlation, which can be related to the transition from 
%  the hard state to a softer  state on a timescale of 2 weeks. The  data 
%  from revolution 56 shows  constant hardness, and are clearly quite different.\\
%Figure~\ref{fig:hard61} focuses on the  hardness evolution during revolution 61 (containing the 
%first flare). We clearly see 2 groups of points, with mean fluxes of $\sim$220 and $\sim$460 mCrab 
%respectively. During the low level flux periods,
% we notice a flux-hardness anti-correlation, which overlaps revolutions
% 57-58 (Figure~\ref{fig:hard54-58}) in the same diagram. 
%The same behavior is reproduced during revolution 63 and is similar to the horizontal branch formed 
% during revolution 56 in Figure~\ref{fig:hard54-58}, suggesting that revolution 56 could
% correspond to flare emission into the hard state (but see below).

\subsection{Spectral modelling}

\begin{figure*}%[!t]
\begin{center}
%\centering
%%\includegraphics[scale=0.7]{figure3a.ps}
\includegraphics[scale=0.5]{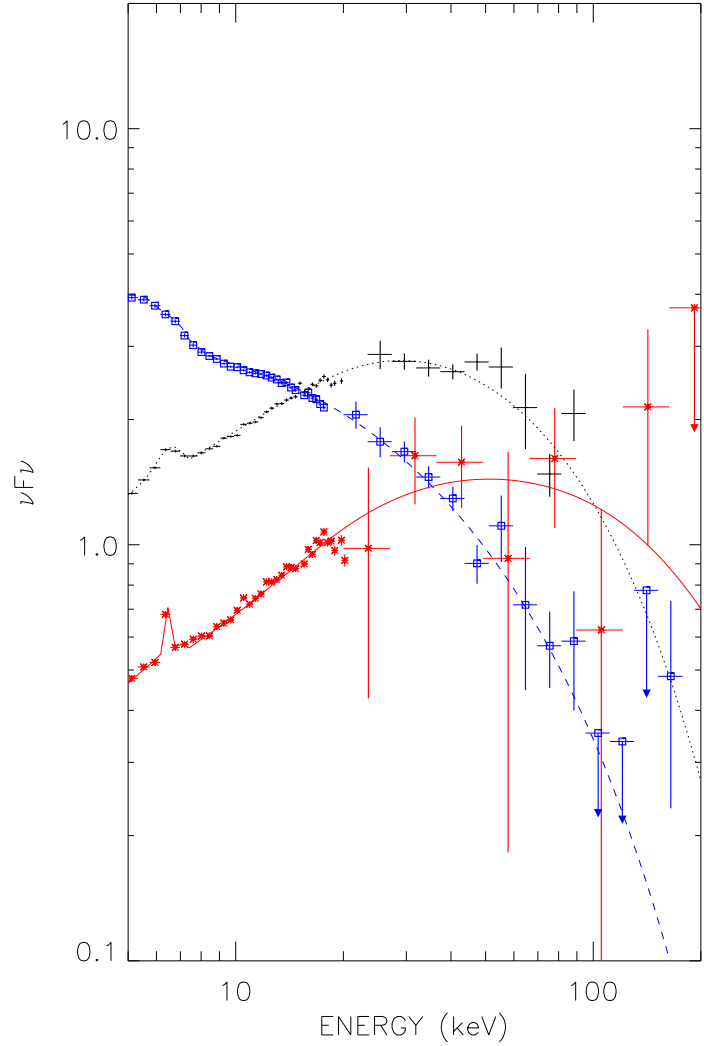}
\end{center}
%\hspace{2cm} 
%\vspace{-0.5cm}
\caption{Spectra from simultaneous PCA/RXTE and SPI/INTEGRAL observations of H1743-322 during revolution 55 (red asterisks)
 and revolution 56 (green points) and during revolution 58 (blue squares). 
 The model used is 
 FN$\times$smedge$\times$phabs$\times$(diskbb+pexrav+gaussian) (see Table 6).}  
\label{fig:spec54-58}
\end{figure*}

Following the temporal evolution described above, we have compared 
the spectra corresponding to revolutions 54-55, 56 and 58 in Fig.~\ref{fig:spec54-58} . For each of them, we have put together 
 the deconvolved SPI and PCA/RXTE spectra.
% , with the 3 available ASM fluxes, converted to the appropriate
%  units through a Crab model (assuming a Crab emission N(E) = 11.6 E$^{-2.09}$ \citep{ASM}).
The spectral evolution is characterized by the peak of the maximum of the energy moving progressively
from 80 keV to a few keV, illustrating the hard to the intermediate state transition.

% From this point of view, the evolution from  revolutions 54-55 to  
%revolution 58 is continuous and the "flare"  seen in revolution 56 could be interpreted in terms 
%of peak emission crossing the 20-36 keV band. The evolution of hardness versus the bolometric
%(2-200 keV) luminosity  (Figure~\ref{fig:lum})  supports such an interpretation, as the
%  evolution of the source hardness is continuous from a hard to a softer state when the luminosity 
%  increases.

\begin{figure*}%[!t]
\begin{center}
\includegraphics[scale=0.5]{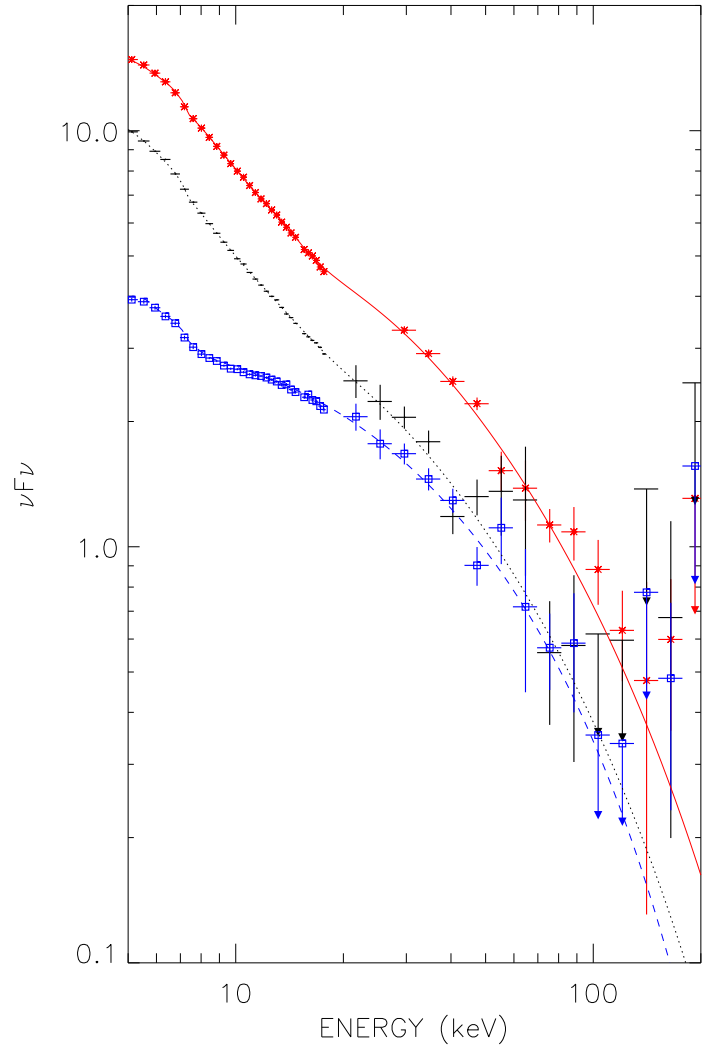}
\end{center}
\caption{Spectra from simultaneous PCA/RXTE and SPI/INTEGRAL observations  of H1743-322 during revolution 58 (blue square)
, revolution 60-63 (no flare events) (red asterisk)
 and revolution 61-63 (flare events) (in green). The model used is 
 FN$\times$smedge$\times$phabs$\times$(diskbb+pexrav+gaussian) (see Table 6).}  
\label{fig:spec58-63}
\end{figure*}

 To study the intermediate and SPL state (revolutions 58 to 63), we separate flare and  no flare emissions,
the flare state being  defined by a 20-36 keV flux greater than 350 mCrab (see Figure~\ref{fig:fig2}).
 Figure~\ref{fig:spec58-63} shows an increase in the X-ray emission alone between 
 revolutions 58 and  60-63, flares excluded, while the flare spectrum is shifted by a factor of 2 above 
  the non-flare spectrum,  keeping roughly the same shape from 5 to 200 keV.

To better quantify the spectral  evolution, we have first fitted PCA and SPI data  separately.

%%%%explanation around RXTE model
We used the standard XSPEC 11.3.1 fitting package to fit the PCA data in a 3-20 keV range.
These data were modelled with several components. In all cases, the spectral 
continuum is described by a multicolor disk black
body (DISKBB in XSPEC) \citep{Mit} plus a powerlaw. We account for interstellar absorption
using PHABS in XSPEC. Individual fits revealed that $N_H$ was consistent
with $2.3\times 10^{22}$ cm$^{-2}$ as determined by Miller el al. (2004).
The $\chi ^2$ was acceptable after including an iron emission
  line modelled by a narrow Gaussian centered and fixed at 6.4 keV.
  A smeared iron edge (Ebisawa et al. 1994) (SMEDGE in XSPEC) -which is kept free- of 6-7 keV
  improves the $\chi ^2$. The smearing width has been always fixed to 10 keV.
 The result of spectral fits in the 3-14 keV band with this model is reported in Table~\ref{tab:table3a}. 
The photon indices increase from 1.4 (rev. 55) up to 2.6 (rev. 60-63) in the PCA/RXTE band while the index
in the 20-300 energy range,
obtained by fitting SPI data by a single powerlaw,
increases from 1.8 to 3.1. This spectral evolution is clearly related to the transition from the 
hard to a softer state.
These values are comparable with the values obtained by HEAO-1 during the 1977-1978 outburst, with the 13-80 keV flux 
varying from 140 to 100 mCrab and the powerlaw index from 2.6 to 2.2 ($\pm$ 0.2) (Cooke et al. 1984; Levine et al. 1984).

 Secondly, we fit the 2 instruments together. A basic cross-calibration of the instruments has
 been performed by comparing the Crab spectrum of both instruments at 30 keV. A normalization factor of
0.96 for the PCA data  relative  to the  SPI data has been determined. However, as the RXTE observations
were never performed during the corresponding SPI observation periods (see Table~\ref{tab:table1}), 
this factor has been kept free. 
Particularly, during the flaring period of revolution 63, RXTE observations 
missed the lowest flux part 
registered during the SPI observations. This explained the low normalization factor 
obtained for the average spectrum of the flare state.

The PCA and SPI spectra have been fitted together using several models. In all
cases, the RXTE spectra have been fitted using the model presented above, the powerlaw component
was replaced first by a powerlaw plus cutoff model (CUTOFFPL in XSPEC), then by COMPTT and PEXRAV model to ajust the high energy emission.
First, we found, using  \textit{ftest} (in XSPEC), that an exponential cutoff is required for 
all revolutions. However, from revolution 60, the 
energy cutoff is poorly constrained (see table~\ref{tab:table3b}). Moreover the powerlaw slope 
was found to be equal to the one 
determined when the RXTE data was fitted separately ($\Gamma _x$=$\Gamma _{\gamma} ^{'}$). 

With the Comptonization model COMPTT (in XPSEC), the
temperature of the disk (T$_{in}$)  in the multicolor disk blackbody model is forced to be equal
 to the soft photon temperature (T$_{o}$) of the Comptonization model.
 We see in Table~\ref{tab:table3c} that the optical depth decreases from about 3.0 down to
 0.3 while the temperature increases from 15 up to $\simeq$ 40 keV.  
 We notice  that the best fit parameters obtained 
for the data set of revolutions 60 to 63 with and without flaring activity are very close.

Then, we used a reflection model (PEXRAV in XSPEC), which is justified by the presence 
of an iron line in all the spectra.
The fit parameters are described in Table~\ref{tab:tableref}.
During the transition hard to intermediate state, the 
energy cutoff increases with 
the powerlaw photon index increases, while a reflection scaling factor of about 0.5 has been found (or fixed) but
is poorly constrained

\section{Discussion}

%H1743-322 presents a temporal and spectral  behaviour similar to that of other 
%black hole candidates. 
%Its rise time (much longer that classical Novae X) and its recurrence time 
%(between 7 and 27 years) make  it nevertheless difficult to classify.
%The Ariel V observation campaign revealed a 2-10 keV flux increase from 500 mCrab in August 1977 
%up to 730 mCrab \citep{Kaluz} in September and a decrease by an order of magnitude six months latter
% \citep{Wo}.
% The source was also observed  during this outburst by HEAO-1 
%  in the 13-80 keV energy range  with a flux of 141 mCrab (August 1977) but still $\sim$ 100 mCrab 
%  in March 1978 \citep{Lev}, and a powerlaw photon index  of 
%  2.6 (+/-0.2) and  2.2 (+/-0.2) respectively\citep{Cook}. The last observation in September 1978 
%  found the source at the 2 $\sigma$ level ($\sim$ 10 mCrab).

\begin{figure*}%[!t]
\includegraphics[scale=0.7]{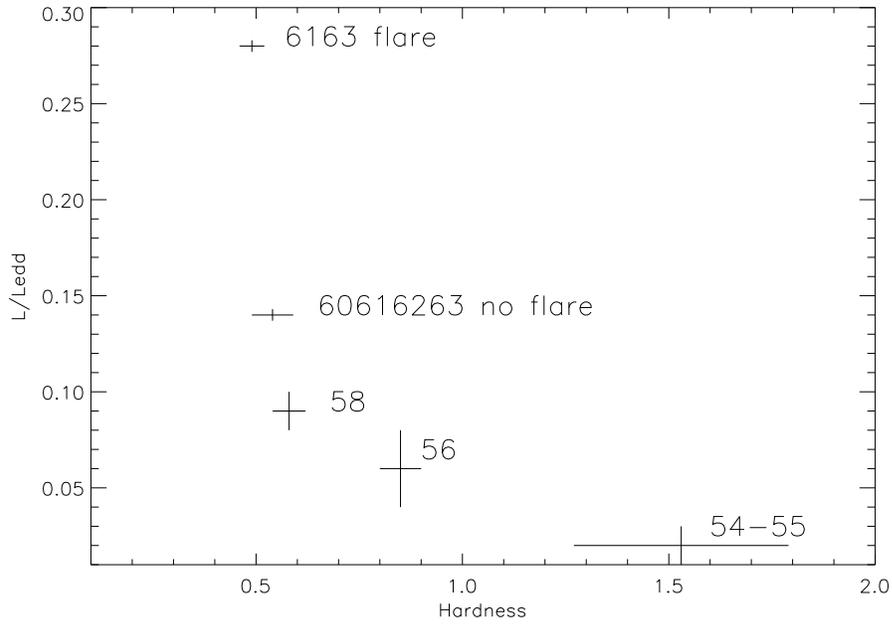}
\caption{Evolution of the bolometric luminosity in Eddington luminosity Ledd as a function of the hardness
in the 23-200 keV energy range. A distance of 8.5 kpc has been assumed.}  
\label{fig:lum}
\end{figure*}

The spectral evolution of the source during the rise to the peak of the outburst can be described in two phases: 
\begin{itemize}
\item[(1)] During the first two weeks, the spectrum undergoes a gradual softening associated with 
  the peak of the SED (spectral energy density) shifting from 80 keV in the hard state down to 10 keV on April 8 (rev. 58) 
   (Figure~\ref{fig:spec54-58}) and leading  to a clear Hardness-Flux anti-correlation (Figure~\ref{fig:hardtot}).
\item[(2)]  Once this softer state is reached the source shows no significant spectral evolution
   despite substantial changes in luminosity. The source exhibits flaring activity 
   on a time scale of 1 day during which, the spectral shape remains  unchanged 
   from soft to hard X-rays while the flux intensity changes by a factor 2-3.
\end{itemize}
This spectral evolution is illustrated in figure~\ref{fig:lum} which displays the evolution of the source in the
bolometric luminosity versus hardness plane. During phase 1 the hardeness decreases 
with little changes in luminosity, 
while during the second phase (after revolution 58) the source exhibit large
 variations in luminosity at nearly constant hardness.

% T
%   The huge flux increase observed between revolutions 54-55 and 56  could suggest that a first 
%  flare is occuring during this hard state period, but we can also 
 % interpret this as the result of a continuous softening of the spectral emission with the peak
  %energy passing through the SPI band during revolution 56.
 %    The observed transition from a  hard to a soft state  can be compared to that of Cyg X-1,
% GX339-4 or other transient sources.
% A geometry proposed by Z02 to describe such an evolution is a hot 
% To test the scenario

 In  Chandra and RXTE 
observations taken in May 2003 \citep{Mil}, the blackbody component dominates the spectral
 emission. The same behaviour is observed in August 2003 in JEM-X  data with a 
 non-detection by IBIS \citep{Cap}.  On the other hand, during our observing period 
 the contribution of the thermal disk component to the 2-20 keV flux 
 remained below 32 \% (see table 4), and well below the 75 \% criterion 
 defining the thermal dominant state (McClintock \& Remillard 2003). 
  Therefore the  source remained in an intermediate State (IS) or in
 a Steep Power Law state  (SPL) (according to the classification of 
 McClintock \& Remillard (2003)) at least until April 22  (MJD 52751) . 

%    with all models. 
% There is no clear evolution of the electron temperature. But above
% revolution 60, it ranges around 30-40 keV with all models.

 %During the initial softening  phase, the electron temperature decreases by a factor of 2 between
 %revolution , and later shows no clear evolution.

In the hard state, the hard X-ray emission is generally believed to be dominated by a hot geometrically
 thick optically thin accretion flow (Shapiro et al. 1976; Narayan \& Yi 1994) surrounded
  by a cold geometrically thin disk (Shakura \& Sunyaev 1973)\footnote{See
  Markoff et al. (2001) and Markoff \& Nowak (2004) for alternative models for the hard state 
  involving a significant fraction of the hard X-ray emission due to the jet}. 
  Phase (1) of the outburst could be due to the gradual decrease of the inner radius of the cold
  accretion disk, associated  with either the cold disk penetrating  the hot inner flow, or the  latter
   collapsing into an optically thick accretion disk with small active regions of hot plasma on top of it 
   \citep{Zdza}. In both cases the enhanced soft photon flux from the disk 
   tends to cool down the hot phase, leading to softer spectra. 
   Our observations tend support this picture:
   our fits indicate that as the hard X-ray spectrum softens, 
   both the inner disk temperature and flux grow progressively, as expected when the 
 accretion disk surface and emission increase  (see Tables~\ref{tab:table3c}).
Although the best fit parameters obtained with the Comptonization model show an increase 
of the hot plasma temperature, the Compton parameter $y=4\tau kT/m_{e}c^2$ decreases by a factor of 5.  
As the $y$ parameter is related to the ratio of Compton luminosity $l_{\rm h}$ (heating of the hot plasma)
 to the soft seed photons luminosity $l_{\rm s}$    ($y\simeq l_{\rm h}/l_{\rm s}$), 
this evolution is consistent  with  an increase of soft photon flux relative to the heating power.
This enhanced cooling can be seen more directly from luminosity measurements, 
by estimating the ratio  $l_{\rm h}/l_{\rm s}$ as $ \Phi_{\rm bol}   / \Phi_{bb}$ 
($ \Phi_{\rm bol}$ is the bolometric luminosity in the 2-200 keV energy range shown in figure 7), which, 
from the numbers shown in Table~\ref{tab:table3c} and in figure 7 decreases
 by a factor of about 5, i.e. roughly consistent with the evolution independently 
 deduced from the Comptonization model.
The observed increase of the hot plasma temperature despite an enhanced soft cooling,
is due to the fact that the optical depth decreased by a larger amount reducing the efficiency 
of the energy transfer from electrons to photons.

The diminishing optical depth could have several causes:
for instance, the presence of a strong hot comptonizing corona is known to be closely associated with 
the steady compact jet of the hard state (Corbel 2004; Fender et al. 2004, hereafter FBG04) and it has been suggested that the hot 
corona forms the base of the jet. 
 The decrease of the coronal optical depth as the source evolves toward the soft state could be related to the
 disappearance of the compact jet. 
 The fact that the transition from phase (1) to phase (2) appears coincident 
 with a major radio outburst, most likely associated with an ejection event \citep{Rupb}, 
 suggests that most of the coronal material could have 
 been wiped out or ejected during this and possibly other less prominent ejections. 
 Alternatively, it is possible that, as the hot plasma condensates into an optically thick disk, 
 the remaining material in the hot corona has a lower density.

During phase (2), the overall geometry would be globally stable despite instabilities 
 leading to chaotic light curves with little spectral variability. Such a strong variability
 could be due to local disk instabilities  
 or strong flares in the corona associated with magnetic reconnection events as those inferred in
  the soft states of Cygnus X-1 (see Zdziarski et al. 2002). Radio emissions lead naturally to
  assume the presence of a jet, at least temporarily. Alternative models including jets should thus be
  considered as they can contribute to hard X-ray emission as already proposed by different
  teams (e.g. Petrucii et al. 2004, Markoff et al. 2001, Georganopoulos et al. 2002).
  %Alternatively, Ferreira et al. (2005) have recently proposed 
  %an accretion-ejection paradigm explaining such a flaring activity in intermediate states.
   %In this scenario, the hard X-ray emission of intermediate states is dominated by ejected electron/positron clouds. 
   %Indeed, the high disk luminosity combined
   %with the presence of a slow MHD jet allows pair creation and acceleration along
    %the jet axis, giving birth to flares and superluminal ejection events. 
   %After rapid expansion,  the same pair blobs are also responsible for the thin radio flare emission 
   %such as that observed during the major radio outburst associated with the transition between phase 1 and phase 2
   %\citep{Rupb}. The fact that, in H1743-322, most of the X-ray flaring activity occurs after 
   %the main radio flare is however not readily explained within this framework.
   
   The presence of a strong radio flare during the state transition seems to be common among
    black hole transients 
 (Corbel 2004; FBG04). 
  FBG04 provides an interesting interpretation: during the initial softening,
   the mildly relativistic jet associated 
  with the canonical  low/hard state persists, but as the disk makes its inward collapse, 
  the jet becomes  unstable and the Lorentz  factor rapidly increases, resulting in an internal shock in 
  the outflow, which is the cause of the strong  optically thin radio emission. 
  
  As we compare the behaviour of H1743-322 with that of
   other transients, it is worth noting that, even though appearing during the softening
   period, near the
   transition from the intermediate to the SPL state, the radio emission occurs during 
   the rising part of the outburst and precedes the
   soft/hard X-ray peak emission. From this point, H1743-322 is clearly atypical.
   %However, the fact that 
   %the radio ejection occured during the initial rise and before the soft/hard peak emission is 
   %quite stricking. 
   From Table 3 in Brocksopp et al. (2002), we see that the radio emission
   occurs after and sometimes (2 cases) simultaneously with the X-ray maximum emission. 
  The flaring activity following the radio emission could be related to the unstable
  state of the jet as proposed by FBG04.

\section{Conclusions}  

The (re)discovery by INTEGRAL of IGR J17464-3213/H1743-322  led to the collection of an important
 amount of data on this source. H1743-322  shows a rather typical behaviour of black
  hole candidates with the presence of various spectral states.
%A huge outburst recorded in X-rays (ASM/RXTE data) is clearly a transition between the 2 states. 
We analysed the SPI/INTEGRAL and  PCA/RXTE observations recorded in March and April 2003, 
which cover
the rising part of the outburst up to the beginning of the maximum of the X-ray emission. 
Initially in a standard hard state, the source spectrum gradually softened until around April 8
 when a major radio flare was reported.  In the framework of the comptonization 
 model, this softening phase can be associated with an optical depth decrease
 (from $\sim$ 3 down to $\sim$ 0.3). 
 After the radio flare, the hard X-ray spectral shape seemed to remain unaffected by a 
 strong (X and $\gamma$-ray) 
 flaring activity. 
 We tentatively identified the softening phase as the intermediate state and the 
 flaring phase as the SPL, and we note that during the outburst of H1743-322 X-ray transient, the end of the softening phase 
  and the optically thin radio outburst are not associated with the peak of 
  the soft/hard X-ray luminosity.

% The absence 
%of any pronounced soft component in the observed spectra suggests that, during the rising
%part of the outburst, the source is in an intermediate state, as  observed for Cyg X-1 (see Fig. 13 from 
% Z02)  and will reach the soft state after the soft X-ray maximum.\\
%We detected two hard X-ray bursts during the intermediate state which are also observed in the
% X-ray band  few days, and take place a few days after a radio flare.

%By the presence of an "off" state, repeated between 7 and 27 years, the weakness of the 
%disk emission during the rise part of the soft
 %state, the presence of flares in the radio and in hard X-rays, H1743-322 shows similarities
 % with both microquasars and GX~339-4. 
 % Further investigations of the temporal an spectral characteristics of the source
 %%should resolve its nature.
%  It is necessary 
% to go more deeply inside the models and test them for a global   scenario explaining all the temporal
%  and spectral caracterictics  of this source.   
%--------------------------------------------------

\acknowledgments
%!!!!!!!!!!!!!!!!!!!!!!!!!!!!!!!!!!!!!!!
The SPI project has been completed under the responsibility and leadership
of the CNES. We are grateful to ASI, CEA, DLR, ESA, INTA, NASA and OSTC
 for support. \\
Specific softwares used for this work have been developed by L. Bouchet.
A.J. acknowledges Emrah Kalemci for its advices concerning RXTE data analyses.
We are grateful to referee for its very fruitful comments, which have clearly improved the paper.
%!!!!!!!!!!!!!!!!!!!!!!!!!!!!!!!

\clearpage


\clearpage

\clearpage

\clearpage

\clearpage

%\begin{table}[htp!]
%\centering
%\begin{tabular}{lccccc} 
% \hline
%Rev. & $t_{start}$ & $t_{stop}$ & $\Delta$t$_{SPI}$ (ks) & $N_p$ &$\Delta$t$_{ASM}$ (ks)\\
%\hline
%\hline
%53& 2003-03-21 02:04:28&2003-03-22 22:48:28 & 23 &20&0.3\\
%54& 2003-03-25 00:28:23& 2003-03-26 15:56:19&  17&15&0.09\\
%55& 2003-03-28 07:44:24& 2003-03-28 16:57:25&   7&5&0.2\\
%56& 2003-03-30 18:02:56& 2003-04-01 15:27:33& 16&12&0.4 \\
%57& 2003-04-03 08:14:02& 2003-04-03 11:18:37 &  4&3&-\\
%58$^*$& 2003-04-06 15:41:03& 2003-04-07 10:29:44 &  23&16& 1.3\\
%59& 2003-04-08 16:19:09& 2003-04-09 07:13:29& 7& 6&0.2\\
%60& 2003-04-12 06:39:25& 2003-04-12 22:04:19& 23 &18& 0.4\\
%61& 2003-04-14 00:20:16& 2003-04-16 14:17:04& 102& 74&2.9\\
%62& 2003-04-18 22:17:24& 2003-04-19 10:46:20& 18&15&0.7\\
%63& 2003-04-20 13:59:37& 2003-04-22 13:58:32 &  66&46&1.7\\
%\hline
%\multicolumn{3}{l}{$*$: radio flare \citep{Rupb}}
%\end{tabular}
%\caption{The INTEGRAL observations of H1743-322 during AO-1:
%2003-03-21 00:00:00 corresponds to IJD 1175 and MJD 52719.
%For each revolution (Rev.), we give the beggining $t_{start}$  and the end $t_{stop}$ of the observation
%and the number of pointings $N_p$. 
%$\Delta$t$_{SPI}$ is the usefull duration for SPI observations. We extracted 
%the ASM/RXTE observations done inside the range time of SPI
%observations and the corresponding observation duration is $\Delta$t$_{ASM}$.} 
%\vspace*{-0.3 cm} 
%\label{tab:table1}
%\end{table}

\begin{table}[htp!]
\centering
\begin{tabular}{lccccccccccc} 
\hline
Rev. & st &SP$_{start}$ &SP$_{stop}$& $\Delta$t$_{sp}$ (ks)&ID&RX$_{start}$& RX$_{stop}$&  Exp.(ks)\\
\hline
\hline
53   &  &52719.53      &52720.49    &   23                  &       &            &             &            \\
54   & &52723.77      &52724.51    &   17.4                &       &            &	     &		  \\
55   & &52726.30      &52726.65    &    7.6                &  80138&52726.80    &52726.87     & 5.7        \\	     
56   & &52729.10      &52729.48    &   16.1                &  80138&52729.79    &52729.83     & 3.6        \\
57   & &52732.27	     &52732.35    &    4.0                &  &            &             &            \\
58$^*$& &52735.71      &52736.42    &   61.2                &  80138&52735.71    &52735.76     &4.4         \\
59   & &52737.63      &52737.92    &    7.0                &  &	       &             &          \\
60   & &52741.21      &52741.80    &   23.2                &  80146& 52741.83   &52741.93     &5.7         \\
61  & nf&52742.93      &52744.6     &    71.9               &  80146&52743.22    &52743.26     &3.7         \\                  
     &  &             &            &                       &  80146&52743.22    &52743.26     &3.7         \\                  
     &  &             &            &           	          &  80146&52744.20    &52744.24     &3.0         \\
61  & f&52744.60      &52744.79    & 8.4                   &       &            &             &3.7         \\                  
  & nf&52744.79      &52745.47    & 21.7                  &  80146&52743.22    &52743.26     &3.7        \\                  
62   & &52747.86      &52748.28    &    17.9               &  80146&52747.61    &52747.68     &5.7        \\
63  &nf &52749.50      &52749.62    &    7.0                &       &            &             &           \\			      	
  & f&52749.71      &52750.78    &    40.1               &  80146&52750.27    &52750.29     & 2.1          \\  
     & &		     &            &                  	  &  80146&52750.31    &52750.36     &3.8      \\
  &nf &52750.78      &52751.18    &    12.5               &  80146&52750.66    &52750.97     &27         \\
     & &              &            &                       &  80146&52751.04    &52751.08     &3.7         \\	
     & &	             &            &                       &  80146&52751.10    &52751.14     &3.1        \\	
  &f &52751.23      &52751.38    &    4.9                &  80146&52751.24    &52751.34     &8.3        \\
     & &	             &            &                       &  80146&52751.37    &52751.40     &2.1        \\
\hline
\multicolumn{5}{l}{$*$: radio flare \citep{Rupb}}
\end{tabular}
\caption{The INTEGRAL observations of H1743-322 during AO-1:
2003-03-21 00:00:00 corresponds to IJD 1175 and MJD 52719.
For each INTEGRAL revolution (Rev.), we give the beginning SP$_{start}$  and the end SP$_{stop}$ 
of the SPI observations in MJD.
$\Delta$t$_{sp}$ is the useful duration for SPI observations. 
The separation between flare (f) and non flare events (nf) for revolutions 61 and 63 has been specified in
column "st".
ID is the identification program number of RXTE observations. RX$_{start}$ and RX$_{stop}$ are
the beginning and the end of RXTE observations taken simultaneously with SPI observations.
Exp. is the exposure time for PCA/RXTE.} 
\vspace*{-0.3 cm} 
\label{tab:table1}
\end{table}

\begin{table}[h]
\centering
\begin{tabular}{lccccc} 
\hline
& Source name  & Flux (mCrab) & Significance& Flux (mCrab) & Significance\\
&              &   20-36 keV  &             &  90-120 keV  &\\
\hline
\hline
1&H1743-322   &  266.98 $\pm$  2.09   &  127.2 &   96.70 $\pm$  7.36 & 13.1\\
2 &1E \textit{region} (a) &  108.31 $\pm$  2.35   &   46.1&   79.00 $\pm$  7.14 & 11.1\\
3   &4U 1700-377   &  158.47 $\pm$  3.30   &   48.0&   55.03 $\pm$  8.83 & 6.2\\
4  & XTE J1720-318 &   39.82 $\pm$  2.34   &   17.0\\
5  & OAO 1657-415  &  134.20 $\pm$  4.21   &   31.9\\
6  & Ginga 1826-24 &   73.23 $\pm$  2.65   &   27.6\\
7   &IGR J17475-2822&   51.17 $\pm$  2.49   &   20.5\\
8   &GX 5-1        &   69.99 $\pm$  2.06   &   33.9\\
9   &GX 354-0      &   60.51 $\pm$  2.36   &   25.6\\
10  &H 1702-429    &   46.12 $\pm$  4.80   &   9.6\\
11 & GX 1+4        &   28.69 $\pm$  2.28   &   12.6\\
12 & GX 3+1        &   32.73 $\pm$  2.36   &   13.9\\
13  &H 1820-303    &   27.62 $\pm$  2.35   &   11.7\\
14  &3A 1822-371   &   30.15 $\pm$  2.50   &   12.0\\
\hline
\multicolumn{5}{l}{note : 1 Crab = 0.19 counts /s in the 20-36 keV energy range}
\end{tabular}
\caption{Catalog of sources surrounding H1743-322 in the 20-36 keV and 90-120 keV energy range. (a) : due to the modest angular
resolution, SPI cannot distinguish all the sources present around 1E 1740.7-2942.} 
\vspace*{-0.3 cm} 
\label{tab:table2}
\end{table}

\begin{table}[h]
\centering
\begin{tabular}{l|cc@{\kern10pt}|@{\kern10pt}ccl} 
\hline
   & \multicolumn{2}{c}{PCA/RXTE}  & \multicolumn{2}{c}{SPI}\\
\hline
Rev.     & $\Gamma  _{x}$ & $\chi$ $^2$(dof) &$\Gamma _{\gamma}$ &$\chi ^2$(dof) \\
\hline
\hline
55       & $1.40^{+0.03}_{-0.13}$& 2.88(17)&$1.8^{+0.2}_{-0.2}$& 0.9(10) \\
56       & $1.33^{+0.17}_{-0.36}$& 2.07(18)&$2.5^{+0.1}_{-0.1}$&2.7(14)  \\
58       & $2.36^{+0.06}_{-0.10}$& 2.61(13)&$2.9^{+0.1}_{-0.1}$&0.9(15)  \\ 
60 to 63 (nf)& $2.65^{+0.03}_{-0.04}$& 1.59(18)&$3.1^{+0.1}_{-0.1}$& 1.2(15)\\                             
61-63    (f) & $2.57^{+0.03}_{-0.03}$& 1.34(17)&$3.0^{+0.1}_{-0.1}$&1.0(14) \\
\hline
\multicolumn{5}{l}{(nf) refers to no flare events and (f) to flare events}
\end{tabular}
\caption{H1743-322 data from PCA and SPI fitted separately.
SPI/INTEGRAL and PCA/RXTE data have been fitted using a powerlaw model  and 
SMEDGE$\times$PHABS$\times$(DISKBB+POWERLAW+GAUSSIAN) (see section 5.4 concerning
the parameters of each components) respectively.
The iron line width was kept free (except for the rev 55,56 and 58:
  it was fixed at 1.0 keV).
  $\Gamma _{x}$ and $\Gamma _{\gamma}$
are the photon powerlaw 
indices  in the 3-14 keV energy band and in the 20-100 keV
energy band respectively.} 
\vspace*{-0.3 cm} 
\label{tab:table3a}
\end{table}

\begin{table}[h]
\centering
\begin{tabular}{l|ccc@{\kern10pt}|@{\kern10pt}ccccl} 
\hline
Rev.     & FN& $\Gamma _{\gamma}$& $\chi ^2$(dof) &  FN& $\Gamma _{\gamma} ^{'}$ & $E _{cut}(keV)$     &$\chi ^2$(dof)\\
\hline
\hline
55          & 1.3& $1.43^{+0.01}_{-0.01}$& 2.8(41)  &  0.8& $1.0^{+0.2}_{-0.2}$ & $46^{+20}_{-11}$ &2.2(40)\\
56          & 0.8& $1.57^{+0.01}_{-0.01}$& 11.0(48) &  0.7& $1.2^{+0.1}_{-0.1}$ & $40^{+6}_{-3}$   &1.6(47)\\
58          & 1.6& $2.47^{+0.01}_{-0.01}$& 3.8(46)  & 1.2& $2.1^{+0.1}_{-0.1}$ & $25^{+2}_{-3}$   &1.4(45)\\ 
60 to 63 (nf)& 0.9& $2.81^{+0.02}_{-0.02}$& 2.7(45)& 0.8& $2.5^{+0.1}_{-0.1}$ & $70^{+56}_{-23}$ &1.9(44)\\                             
61-63    (f) & 0.7&$2.76^{+0.02}_{-0.02}$& 3.2(43)& 0.7& $2.6^{+0.1}_{-0.1}$ & $103^{+25}_{-29}$&1.8(42)\\
\hline
\multicolumn{5}{l}{(nf) refers to no flare events and (f) to flare events}
\end{tabular}
\caption{H1743-322 data from PCA and SPI fitted jointly using :
FN$\times$SMEDGE$\times$PHABS$\times$(DISKBB+POWERLAW+GAUSSIAN) (see section 5.4 concerning
the parameters of each components). FN is the normalization factor between SPI/INTEGRAL and PCA/RXTE.
$\Gamma _{\gamma}$ is the photon powerlaw index.
POWERLAW was replaced by a cutoff powerlaw model : 
$\Gamma _{\gamma} ^{'}$ and $E_{cut}$ are the photon powerlaw index and the energy cutoff respectively.
The iron line width was kept free (except for the rev 55,56 and 58:
  it was fixed at 1.0 keV).} 
\vspace*{-0.3 cm} 
\label{tab:table3b}
\end{table}

\begin{table}[h]
\renewcommand{\arraystretch}{0.01}
%\centering
\begin{tabular}{lccccccccc} 
\hline
rev   &  T$_{in}$   &$\tau$  &kT    & $\Phi _{\gamma}$ $\times 10^{-9}$ &$\Phi _{bb}$ $\times 10^{-9}$& $\Phi _{disc}$&$\textit{EW}$& FN &$\chi ^2$(dof)\\	
      &   {\footnotesize keV}       &	     & {\footnotesize keV }  &  ERG   &  ERG            &       {\footnotesize \%}                     &  {\footnotesize eV} &       &      \\
\hline
\hline
55	&$0.27^{+0.02}_{-0.03}$ &$3.13^{+0.03}_{-0.03}$   &$17^{+3}_{-5}$    & 0.08  & 0.2 &2&74  & 0.9  &1.99(37)\\
56	&$0.41^{+0.02}_{-0.01}$ &$3.39^{+0.03}_{-0.09}$   &$15^{+1}_{-1}$    & 0.05  & 0.4 &2&147 & 0.9  &1.58(48)\\
58	&$0.39^{+0.10}_{-0.02}$ &$1.00^{+0.10}_{-0.11}$   &$22^{+5}_{-5}$    & 1.58  & 5.9 &7&245 & 1.5  &0.74(50)\\
6061 nf	&$0.62^{+0.04}_{-0.04}$ &$0.25^{+0.03}_{-0.03}$   &$38^{+2}_{-2}$    & 1.02  & 7.2 &7&143 & 0.7  &2.00(44)\\
6163 fl	&$0.65^{+0.03}_{-0.01}$ &$0.26^{+0.03}_{-0.02}$   &$37^{+2}_{-2}$    & 4.89  & 6.7 &32&173& 0.5  &2.57(41)\\  
\hline
\multicolumn{8}{l}{(nf) refers to no flare events and (f) to flare events, ERG = ergs cm$^{-2}$ s$^{-1}$}
%\multicolumn{5}{l}{ERG = ergs cm$^{-2}$ s$^{-1}$}
\end{tabular}
\caption{Simultaneous PCA-SPI fit with a Comptonization model. The model used was 
described by the following XSPEC components (explained in section 5.4): 
FN$\times$SMEDGE$\times$PHABS$\times$(DISKBB+COMPTT+GAUSSIAN). 
T$_{in}$ 
  is the inner disk temperature,
$\tau$ the optical depth and kT the plasma temperature. 
$\Phi _{\gamma}$ and $\Phi _{bb}$ (with an error of 10 \%) are the flux in the 20-100 keV 
  energy range and the
  flux of the disk 
  blackbody component in the 2-20 keV energy range respectively. $\Phi _{disc}$ is the disc-flux fraction in the
  2-20 keV energy range.
  The iron line width was kept free (except for the rev. 61,63 during the flaring activity :
  it was fixed at 0.1 keV) and $\textit{EW}$ is the equivalent width. 
  FN is the normalization factor between SPI/INTEGRAL and PCA/RXTE.}
\vspace*{-0.3 cm} 
\label{tab:table3c}
\end{table} 

\begin{table}[h]
\renewcommand{\arraystretch}{0.01}
\centering
\begin{tabular}{lccccccc} 
\hline
Rev.    &T$_{in}$          &	   $\Gamma$   &	   $E_{cut}$	  &   $\Omega /2 \pi$       &     FN	&   $\chi ^2$(dof)\\
	&keV&              &                  keV&                        &       &   \\
\hline
\hline
55	&$0.29^{+0.36}_{-0.29}$&$1.07^{+0.15}_{-0.19}$& $80^{+145}_{-37}$ & 0.5 fr	           & 1.0 & 1.47(40)\\
56	&$0.34^{+0.46}_{-0.08}$&$1.45^{+0.12}_{-0.18}$&	$62^{+13}_{-21}$  & $0.58^{+0.21}_{-0.32}$ & 1.0  & 1.30(46)\\
58	&$0.37^{+0.31}_{-0.06}$&$2.37^{+0.06}_{-0.05}$&	$71^{+26}_{-17}$  & $0.43^{+0.25}_{-0.12}$ & 1.3  & 0.80(45)\\
6063 nf	&$1.33^{+0.03}_{-0.03}$&$2.58^{+0.10}_{-0.15}$&	$88^{+85}_{-37}$ & $0.49^{+0.26}_{-0.29}$& 0.8 & 1.70(42)\\
6163 f	&$1.57^{+0.01}_{-0.01}$&$2.58^{+0.03}_{-0.03}$&	$99^{+28}_{-19}$ & 0.5 fr.& 0.8  & 2.39(39)\\
\hline
\multicolumn{5}{l}{(nf) refers to no flare events and (f) to flare events}
\end{tabular}
\caption{Simultaneous PCA-SPI fit with the Reflection model. The model used was 
described by the following XSPEC components (explained in section 5.4): 
FN$\times$SMEDGE$\times$PHABS$\times$(DISKBB+PEXRAV+GAUSSIAN).
T$_{in}$ 
  is the inner disk temperature, $\Gamma$ the powerlaw photon index and $E_{cut}$ the energy cutoff.
$\Omega /2 \pi$ is the reflection scaling factor.
 The iron line width was kept free (except for the rev. 61,63 during the flaring activity :
  it was fixed at 0.1 keV). FN is the normalization factor between SPI/INTEGRAL and PCA/RXTE.}
\vspace*{-0.3 cm} 
\label{tab:tableref}
\end{table}

\end{document}